\documentclass[conference]{IEEEtran}
\usepackage{slashbox}
\usepackage{amsmath}    % need for subequations
\usepackage{graphicx}   % need for figures
\usepackage{graphics}   % need for figures
\usepackage{verbatim}   % useful for program listings
\usepackage{color}      % use if color is used in text
\usepackage{subfigure}  % use for side-by-side figures
\usepackage{array}
\begin{document}

\title{Improved Quantum LDPC Decoding Strategies for the Misidentified Quantum Depolarization Channel}

% author names and affiliations
% use a multiple column layout for up to three different
% affiliations

\author{\IEEEauthorblockN{Yixuan Xie, Jun Li, Robert Malaney and Jinhong Yuan}\\
\IEEEauthorblockA{School of Electrical Engineering and Telecommunications\\
The University of New South Wales, Sydney, Australia\\
%Sydney, Australia \\
Email: Yixuan.Xie@student.unsw.edu.au, Jun.Li@unsw.edu.au,
R.Malaney@unsw.edu.au, J.Yuan@unsw.edu.au}} \maketitle
% make the title area
\maketitle

\begin{abstract}

%In this work we probe the impact of channel estimation on the performance of quantum LDPC codes.  Our channel estimation is based on an optimal estimate of the relevant decoherence parameter via  its quantum Fisher information. Using state-of-the art quantum LDPC codes designed for the  quantum depolarization channel, and utilizing  various quantum probes with different entanglement properties, we show how the performance of such codes can deteriorate by an order of magnitude when optimal channel identification is fed into a belief propagation decoding algorithm. Our work highlights the importance in quantum communications of a viable channel identification campaign prior to decoding, and highlights the trade-off between entanglement  consumption and quantum LDPC code performance.
Quantum cryptography via key distribution mechanisms that utilize quantum entanglement between sender-receiver pairs will form the basis of future large-scale quantum networks. A key engineering challenge in such networks will be the ability to correct for decoherence effects in the distributed entanglement resources. It is widely believed that sophisticated quantum error correction codes, such as quantum low-density parity-check (LDPC) codes, will be pivotal in such a role. However, recently the importance of the channel mismatch effect in degrading the performance of deployed quantum LDPC codes has  been pointed out.
%It was found that an order of magnitude degradation in the qubit error performance was found even if optimal information on the channel identification was assumed. However, although such previous studies indicated the level of degradation in performance, no alternate decoding strategies had been proposed in order to reduce the degradation.
In this work we help remedy this situation by proposing new quantum LDPC decoding strategies that can significantly reduce performance degradation by as much as $50\%$. Our new strategies for the quantum LDPC decoder are based on previous insights from classical LDPC decoders in mismatched channels, where an asymmetry in performance is known as a function of the estimated channel noise. We show how similar asymmetries  carry over to the quantum depolarizing channel, and how an estimate of the depolarization flip parameter weighted to larger values leads to significant performance improvement.

\end{abstract}

\IEEEpeerreviewmaketitle

\hspace{0.3cm}
\section{Introduction}

Quantum cryptography  is considered to be one of the main applications that will be carried over emerging large-scale quantum communication networks. Although the original quantum key distribution  (QKD) protocol  \cite{bb84}, has been the first to be commercialized, it is widely believed that future networks will run commercialized versions of QKD based on the quantum entanglement \cite{ekert}. Indeed,  distributed pairs of entangled qubits (or qudits) will be the backbone of such  future quantum networks, forming the key resource used in almost all quantum communication applications. A key engineering challenge will be the protection of these distributed entangled resources from ongoing decoherence effects, through entanglement distillation or quantum error correction  (distillation can be viewed as a form of error correction). The use of state-of-the-art quantum low-density parity-check (LDPC) codes in this regard has been the subject of much focus (\emph{e.g.} \cite{chau}).

The existence of quantum error correction codes (QECC) was initially shown by Shor \cite{Shor:1995},  Calderbank \cite{Calderbank1996} and Steane \cite{Steane}, with generalization to stabilizer codes shown by Gottesman \cite{Gottesman1996}. These works, amongst others, outlined the relationships  quantum error-correction codes  have to classical codes, leading to a pathway for the most successful classical codes, \emph{e.g.} classical LDPC codes, to be readily converted to quantum codes. A more detailed history on the development of QECC can be found elsewhere \emph{e.g.} \cite{Nielsen:2000}, \cite{Knill1997}. Quantum LDPC codes based on finite geometry were first proposed in \cite{Postol}, followed by the \emph{bicycle} codes proposed in \cite{MacKay2004}.   More recently, many works attempting to improve quantum LDPC code performance have been published, e.g. \cite{Tan:2010} \cite{Hagiwara:2011}, and \cite{kenta2011} based on quasi-cyclic structures  (such structures reduces the complexity of encoding and decoding).

Classical LDPC codes were originally proposed by Gallager in his thesis in 1961 \cite{Gallager:19633}, however, LDPC codes remained largely unnoticed
%(although see \cite{lpc1}, \cite{lpc2}, \cite{lpc3})
until their re-discovery in the mid '90s \cite{mack1} \cite{MN1996}.  Since then many hundreds of papers have been published outlining the near optimal performance of LDPC codes over a wide range of noisy wireless communication channels. In almost all of such previous works it was assumed that the characteristics of the noisy wireless channel was known. However, the  reality is that in many cases an exact determination of the wireless channel  is unavailable. Indeed,  several works have in fact  investigated  the case where a channel mismatch (or channel misidentification) occurs, which in turn impacts on the performance of the LDPC  decoder (e.g. \cite {mackay1997}).
%\cite{Lui}

From the perspective of the work reported on here, the most interesting aspect of such channel mismatch studies is the asymmetry in the LDPC code performance as a function the channel crossover probability for the binary symmetric channel (BSC). In fact, the main focus of the work described here is an investigation of whether such asymmetric LDPC  code performance carries over from the classical BSC to quantum LDPC codes operating over the quantum depolarizing channel.

Recently in \cite{ShaneACTW2012} the impact of channel mismatch effects on the performance of quantum LDPC codes over the quantum depolarizing channel was highlighted. In previous investigations of the performance of quantum LDPC codes it had been assumed that perfect knowledge of the quantum channel exists. Of course in practice this is not the case.
%In  \cite{ShaneACTW2012}, even though  optimal estimates of the quantum depolarizing channel were used, the remaining channel mismatch still led to  an  order-of-magnitude degradation in the quantum sum-product LDPC decoder performance.
In this paper, we further investigate the behavior and the robustness of the sum-product decoding algorithm  over the quantum depolarizing channel. Interestingly, an asymmetry behavior in performance is observed as a function of the estimated channel flip probability, showing that the performance of a quantum LDPC code would experience a reduced degradation when the channel is overestimated instead of underestimated, provided the overestimated channel knowledge still within the threshold limit of the code. Based on these observations, a new decoding strategy is proposed that can improve  quantum codes performance by as much as $50\%$.

In section II we discuss the behavior of the classical sum-product decoder under channel mismatch conditions. In section III we  briefly review quantum communications and the \textit{stabilizer} formalism for describing QECCs, and discuss their relationship to classical codes. In section III we also explore the behavior of a quantum decoder when simulating over a quantum depolarizing channel and show how the decoding strategy we outline here leads to a significant improvement in performance relative to decoders that simply utilize the estimated channel parameter. Lastly, in section IV we draw some conclusions and discuss future works.

\hspace{0.3cm}

\section{Behavior of Classical Sum-Product Decoder}

It is well known in classical coding that low-density parity-check codes are good rate achievable codes \cite{Gallager:19633} \cite{MN1996}, given an optimal decoder. The best algorithm known to decode them is the sum-product algorithm, also known as iterative probabilistic decoding or belief propagation (BP). The performance of sparse-graph codes can be improved if knowledge about the channel is known at the decoder side. However, in practical situations the decoder unlikely to know the channel's characteristics exactly, thus, the robustness of the decoder to channel mismatches is also an important issue when designing practical codes.

\begin{figure}[htp]
\centering
\includegraphics[width=3.5in]{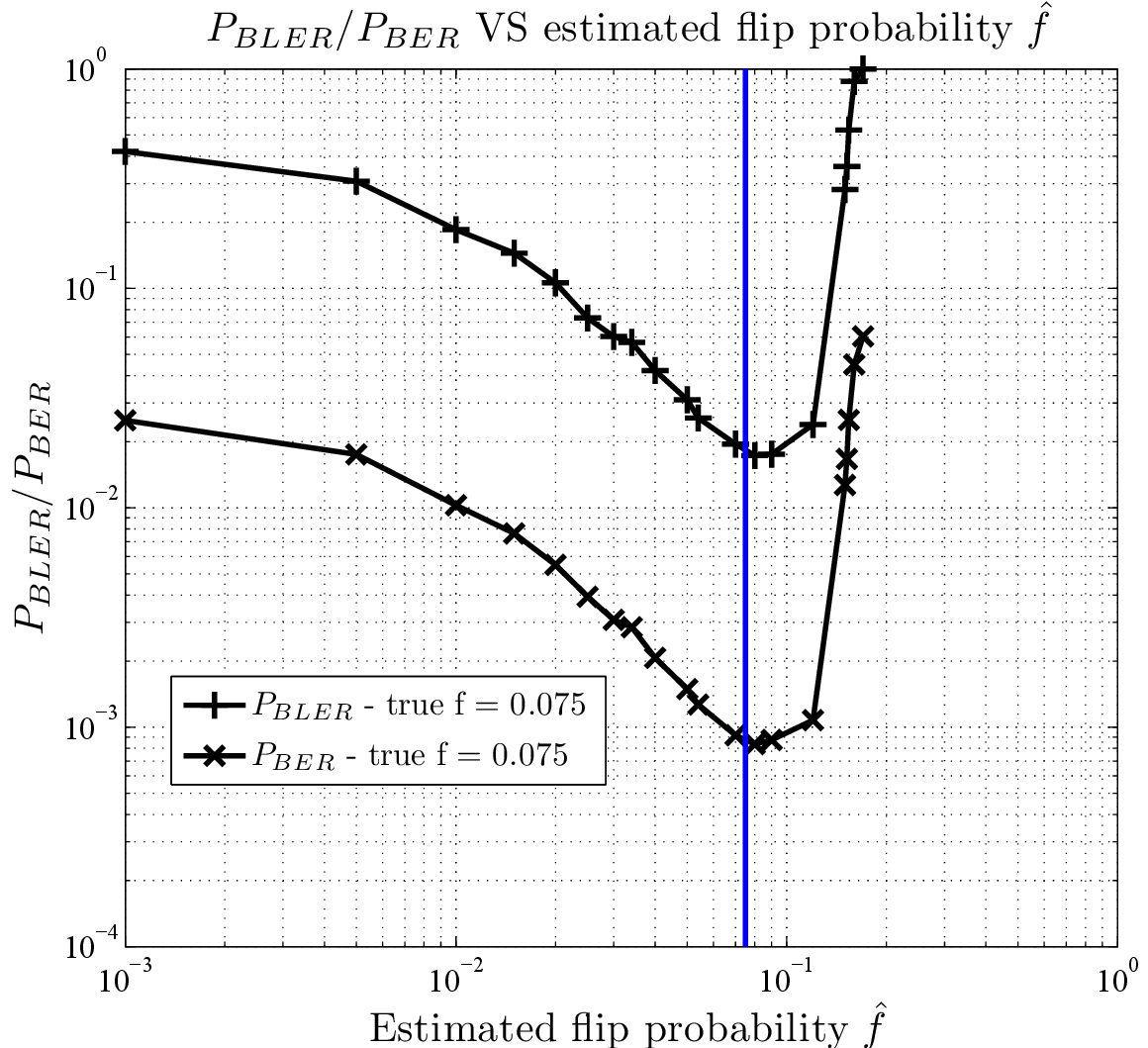}
\vspace{-0.3cm}
\caption{\hspace{0.1cm}Probability of block error as a function of estimated flip probability when the true flip probability is fixed.}
\label{fig:classicalbehavior}
\end{figure}

In \cite{mackay1997}, MacKay \emph{et.al} investigated the sensitivity of Gallager's codes \cite{Gallager:19633} to the assumed noise level (classical bit-flip probability) when decoded by belief propagation. A useful result therein is that the belief propagation decoder for LDPC codes appears to be robust to channel mismatches because the block error probability is not a very sensitive function of the assumed noise level. In addition, an underestimation of channel characteristics deteriorates the performance more compared to an overestimation of channel characteristics. This behavior is shown in Fig. \ref{fig:classicalbehavior}.

Our results shown in Fig. \ref{fig:classicalbehavior} are for a rate half code of block length $N = 2040$ over a binary symmetric channel. The code is a $(3,6)$ regular LDPC codes which is constructed with the length of the cycle maximized. The plot shows the probability of block and bit error $(P_{BLER}/P_{BER})$ as a function of assumed flip probability $\hat{f}$ when the true flip probability $f$ is fixed throughout the simulation.

By inspection, the plotted result shows a similar behavior to that found by MacKay in \cite{mackay1997}.
The vertical straight line indicates the true value of the noise level, and the minimum point of the plot
is approximately at the intersection between the lines. This infers that an optimal performance of a
practical sum-product decoder can be achieved when the input of decoder is the true noise level (true flip
probability). The slope towards the left of the graph is  steeper than the slope towards the right,
indicating that underestimation of the noise level degrades the performance more so than overestimation does.
However, when the estimated noise level is far too large, there is a significant increase in the error
probability. Such higher flip probabilities can be thought of as the classical Shannon's limit (in this case,
the Shannon's limit for rate $1/2$ code is $0.11$), which theoretically represents the threshold $(f_{thr})$ for the
noise level that guarantees reliable transmission at a certain rate.

%\subsection{CSS Codes}
%As mentioned earlier, an important class of codes are the \textit{CSS Codes}  \cite{Calderbank1996}\cite{Steane}. These have the form
%
%\begin{equation}
%\label{CSS}
%A = \left(\begin{array}{c|c}
%\begin{array}{c}
%H \\
%0 \\
%\end{array}
%&
%\begin{array}{c}
%0 \\
%G \\
%\end{array}
%\end{array}
%\right)
%\end{equation}
%where $H$ and $G$ are $M_{H}\times N$ and $M_{G}\times N$ matrices, respectively,  ($M_{H}$ does not necessary equal to $M_{G}$). Requiring $HG^{T} = 0$ ensures that constraint (\ref{binaryConstraint}) is satisfied.

%If $G = H$, the resulting CSS code structure is called a \textit{dual-containing code}. Most classical (good) LDPC codes do not satisfy the constraint (\ref{binaryConstraint}).

\section{Simulations for Improved Channel Decoding}
Motivated by the decoding asymmetry discussed above for classical LDPC codes, we now wish to explore whether a similar asymmetry in decoding performance is achieved for quantum LDPC codes. As stated below several well-known classes of quantum codes, such as quantum stabilizer codes  can be designed from existing classical codes. Upon construction of such codes we will then investigate the decoding performance under asymmetrical estimates of the quantum channel parameters. The quantum channel we investigate is the widely adopted depolarization channel.

\subsection{Quantum LDPC Codes}

%\subsection{Preliminary on Quantum Stabilizer Codes}
\label{DefStabiler}
A stabilizer generator $S$ that encodes $K$ qubits in $N$ qubits consists of a set of Pauli operators on the $N$ qubits closed under multiplication, with the property that any two operators in the set \textit{commute}, so that every stabilizer can be measured simultaneously.
%An example of a stabilizer generator $S$ is shown below for $K=1, N=5$ representing a rate $\frac{1}{5}$ quantum stabilizer code,
%\begin{equation}
%\label{stabilizer}
%S = \left(\begin{array}{ccccc}
%Z & Z & X & I & X \\
%X & Z & Z & X & I \\
%I & X & Z & Z & X \\
%X & I & X & Z & Z
%\end{array}\right).
%\end{equation}

Consider now a set of error operators $\left\{E_{\alpha}\right\}$ taking a state $\left|\psi\right\rangle$ to the corrupted state $E_{\alpha}\left|\psi\right\rangle$. A given error operator either commutes or anti-commutes with each stabilizer $S_{i}$ (row of the generator $S$) where $i=1 \ldots N-K$. If the error operator commutes with $S_{i}$ then
\begin{equation}
\label{errorcommu}
S_{i}E_{\alpha}\left|\psi\right\rangle = E_{\alpha}S_{i}\left|\psi\right\rangle = E_{\alpha}\left|\psi\right\rangle
\end{equation}
and therefore $E_{\alpha}\left|\psi\right\rangle$ is a $+1$ eigenstate of $S_{i}$. Similarly, if it anti-commutes with $S_{i}$, the eigenstate is $-1.$ The measurement outcome of $E_{\alpha}\left|\psi\right\rangle$ is known as the \textit{syndrome}.

%\subsection{Conversion between Quantum and Classical Codes}

To connect quantum stabilizer codes with classical LDPC codes it is useful to describe any given Pauli operator on $N$ qubits as a product of an $X$-containing operator, a $Z$-containing operator and a phase factor $\left(+1,-1,i,-i\right)$.
%For example, the first row of matrix (\ref{stabilizer}) can be expressed as
%
%\begin{align}
%ZZXIX = (IIXIX) \times (ZZIII).
%\end{align}
Thus, we can directly express the $X$-containing operator and $Z$-containing operator as separate binary strings of length $N$. In the  $X$-containing operator a $`1'$ represents the  $X$ operator (likewise for the $Z$ operator), and $0$ for $I$. The resulting binary formalism of the stabilizer is a matrix $ A = \left(A_{1} | A_{2}\right)$ of $2N$ columns and $M = N-K$ rows,  where $A_{1}$ and $A_{2}$ represent $X$-containing and $Z$-containing operators, respectively
%\emph{Example 1:} For example, the set of stabilizers in (\ref{stabilizer}) appears as the binary matrix $A$
%
%\begin{equation}
%\centering
%\label{binaryStabilizer}
%A = \left(A_{1} | A_{2}\right) = \left(\begin{array}{c|c}
%\begin{array}{ccccc}
%0 & 0 &  1 & 0 & 1 \\
%1 & 0 &  0 & 1 & 0 \\
%0 & 1 &  0 & 0 & 1 \\
%1 & 0 &  1 & 0 & 0 \\
%\end{array}
%&
%\begin{array}{ccccc}
%1 &  1 & 0 & 0 & 0 \\
%0 &  1 & 1 & 0 & 0\\
%0 &  0 & 1 & 1 & 0\\
%0 &  0 & 0 & 1 & 1\\
%\end{array}
%\end{array}
%\right).
%\end{equation}

Due to the requirement that stabilizers must commute, a constraint  on a general matrix $A$ can be written as  \emph{e.g.} \cite{MacKay2004}.

\begin{equation}
\centering
\label{binaryConstraint}
A_{1}A_{2}^{T}+A_{2}A_{1}^{T} = 0.
\end{equation}

To summarize, the property of stabilizer codes can be directly inferred from classical codes. Any binary parity-check matrix of size $M\times 2N$ that satisfies the constraint in (\ref{binaryConstraint}) defines a quantum stabilizer code with rate $R = \frac{K}{N}$ that encodes $K$ qubits into $N$ qubits.

\subsection{Quantum Channel Estimation}
The issue of quantum channel identification  (quantum process tomography) is of fundamental importance  for a range of practical quantum information processing problems (\emph{e.g.}  \cite{Nielsen:2000}). In the context of LDPC quantum error correction codes, it is normally assumed that the quantum channel is known perfectly in order for the code design to proceed. In reality of course, perfect knowledge of the quantum channel is not available - only some estimate of the channel is available. The quantum depolarization channel of some states can be defined as $\varepsilon ({\rho _s}) = (1 - f){\rho _s} + \frac{{f}}{3}\sum\limits_{j = 1}^3 {{\sigma _j}} {\rho _s}{\sigma _j}$, where $f$ is the \emph{true flip probability}.\footnote{Note although same symbol use in section II, its new usage here should be clear from the context.}
%To make progress we will assume a depolarization channel with some parameter $f_d$.

%Given some initial system state $\left| {{\Psi _s}} \right\rangle $, a decoherence model can be built by studying the time evolution of the system state's interaction with some external environment. In terms of the density operator ${\rho _s} = \left| {{\Psi _s}} \right\rangle \left\langle {{\Psi _s}} \right|$,  the evolution of ${\rho _s}$ in a channel, which can be written as $\varepsilon \left( {{\rho _s}} \right)$, is a completely positive, trace preserving, map which provides the required evolution of ${\rho _s}$. The depolarization parameter, $f_d$, of a qubit where $0 \le f_d \le 1$, is defined such that $f_d=1$ means complete depolarization and  $f_d=0$ means no depolarization.
% In terms of the well-known Pauli matrices ${\sigma _i}$ (here $i=0,1,2,3)$,
%the depolarization channel for a single qubit can be defined as $\varepsilon \left( {{\rho _s}} \right) = (1 - f_d){\rho _s} + f_d\frac{{{\sigma _o}}}{2}$. Note that it is also possible to parameterize

In what follows  we will assume the true value of $f$ is unknown \emph{a priori}, and must first be measured via some channel identification procedure. This estimate of $f$, which we will refer to as $\hat f$, will be used in a decoder in order to measure its performance relative to a decoder in which the true $f$ is utilized.

In general, quantum channel identification proceeds by inputting a known quantum state $\sigma$ (the probe) into a quantum channel $\Gamma_p$ that is dependent on some parameter $p$ (in our case $p=f$). By taking some quantum measurements on the output quantum state $\Gamma_p(\sigma)$ which leads to some result $R$, we then hope to estimate $p$. The input quantum state may be unentangled, entangled with an ancilla qubit (or qudit), or entangled with another probe. Multiple probes could be used, or the same probe can be recycled (\emph{i.e}. sent through the channel again).

As can be imagined many experimental schemes could be developed along these lines, and the performance of each scheme (\emph{i.e.} how well it estimates the true value of the parameter $p$) could be analyzed. However, in this study we will take a different tact. Here we will simply assume an experimental set-up is realized that obtains the information-theoretical \emph{optimal} performance.

Optimal channel estimation via the use of the quantum Fisher information has been well studied in recent years, particularly in regard to the determination of the parameter $f$ of the depolarizing channel  (\emph{e.g.} \cite{Fujiwara2001}, \cite{Sasaki2002}, \cite{Fujiwara2003}, \cite{Frey2010}, \cite{Frey2011}). Defining ${\rho _f} = {\Gamma _f}(\sigma )$, the
 quantum Fisher information about $f$ can be written as
 \begin{align*}
 J(f) = J\left( {{\rho _f}} \right) = {\rm{tr}}\left[ {{\rho _f}} \right]L_f^2,
 \end{align*}
where ${L_f}$ is the symmetric logarithmic derivative defined implicitly by
\begin{align*}
 2{\partial _f}{\rho _f} = {L_f}{\rho _f} + {\rho _f}{L_f},
\end{align*}
and where  ${\partial _f}$ signifies partial differential w.r.t. $f$. With the quantum Fisher information in hand, the quantum Cramer-Rao bound can then be written as
 \[{\rm{mse}}\left[ {\hat f} \right] \ge {\left( {N_mJ(f)} \right)^{ - 1}}\]
where ${\rm{mse}}\left[ {\hat f} \right]$
is the mean square error of the unbiased estimator ${\hat f}$, and $N_m$ is the number of independent quantum measurements.
%In the simulations pursued here we will assume the channel is constant over the block length of the codeword, and unless otherwise stated we assume $N_m=1$. Further, we will assume two different cases for the quantum probe. In case $A$ we will assume the qubit probe is in a pure unentangled state, and as such \cite{Fujiwara2001} the quantum Fisher information about $f$ relevant to each codeword can be shown to be  $J\left( f \right) = {\left[ {f\left( {2 - f} \right)} \right]^{ - 1}}$. In case $B$ we adopt the scenario where one pair of maximally entangled qubit pairs is consumed per transmission of each codeword (one of the qubits traverses the channel). In this latter case the quantum Fisher information about $f$ relevant to each codeword can be shown to be
%$J\left( f \right) = {\left[ {f\left( {\frac{4}{3} - f} \right)} \right]^{ - 1}}$ \cite{Fujiwara2001}. Similar expressions for qudit probes are available \cite{Frey2011}.

The performance results in \cite{ShaneACTW2012} are obtained by randomly estimating a flip probability from a truncated normal distribution at the decoder side, given the mean square error of the unbiased estimation  ${\hat f}$.
%In return, the performance is degraded approximately an order of magnitude.

\subsection{Quantum Decoding Algorithm}

%In quantum communications system, the quantum channel output can not be realized directly. Instead, the channel output is \emph{measure} to provide binary results in order to visualize the output without destroy the encoded quantum states. The appropriate decoding algorithm for quantum LDPC codes is the classical sum-product algorithm, where the input of the decoding algorithm is the syndrome.

The appropriate decoding algorithm to decode quantum LDPC codes is based on the classical sum-product algorithm since the most common quantum channel model, namely the quantum depolarizing channel, is analogous to the classical $4$-ary symmetric channel. The received values at the decoder side can be mapped to measurement outcomes $s \in \left\{1, -1\right\}^{M}$ (syndrome) of the received qubit sequence, and this syndrome is then used in error estimation and recovery. Assuming an initial quantum state representing a codeword, the initial probabilities $p_i$ for the $ith$ qubit of the state undergoing an $X$, $Y$ or $Z$ error are

\begin{equation}
\label{LLRBSC}
p_{i} = \left\{ {\begin{array}{*{20}{c}}
{{f}}&{for}&{X, \  Y, \ and  \ Z}\\
{1 - {f}}&{for}&{I}
\end{array}} \right.,
\end{equation}
where $f$ is the flip probability known at the decoder.

The standard BP algorithm operates by sending messages along the edges of the Tanner graph. Let $u_{b_{i} \to c_{j}}$ and $u_{c_{j} \to b_{i}}$ denote the messages sent from bit node \textit{i} to check node \textit{j} and messages sent from check node \textit{j} to bit node \textit{i}, respectively. Also denote $\textit{N}(b_{i})$ as the number of neighbors of bit node $i$, and define $\textit{N}(c_{j})$ as the number of neighbors of check node $j$.

To initialize our algorithm, each qubit node sends out a message to all its neighbors equal to its initial probability value $p$ obtained according to equation (\ref{LLRBSC}). Upon reception of these messages, each check node sends out a message to its neighboring qubit node given by
\begin{equation}
\label{checktobit}
u_{{c_j} \to {b_i}} = \sum\limits_{{t_{1 \ldots n}} \in \{ t |t \  \circ \ {c_j}^T = {s_j})\} } \ \ {\prod\limits_{b_i' \in N({c_j})\backslash {b_i}} {u_{{b_{i'}} \to {c_j}}}}
%p({r_i}|{t_i})}
\end{equation}
 where $N\left(c_{j}\right)\backslash b_{i}$ denotes all neighbors of check node $j$ except qubit node $i$, and the summation is over all possible error sequences $t_{ 1\ldots N}$. Each bit node then sends out a message to its neighboring checks given by

\begin{equation}
\label{bittocheck}
u_{{b_i} \to {c_j}} = p_i\prod\limits_{{c_{j'}} \in N({b_i})\backslash {c_j}} {u_{{c_{j'}} \to {b_i}}}
\end{equation}
where $N\left(b_{i}\right)\backslash c_{j}$ denotes all neighbors of bit node $i$ except check node $j$. Equations (\ref{checktobit}) and (\ref{bittocheck}) operate iteratively until the message is correctly decoded or the maximum pre-determined iteration number is reached. The decoder outputs an tentative decision when a valid sequence of $t_{1\hdots N}$ has the same syndrome as $(s_{1},\hdots s_{M})$.

%The beliefs for each qubit node are computed from all the incoming messages as follows
%\begin{equation}
%\label{bittocheck}
%u_{{b_i} \to {c_j}} = p_i\prod\limits_{{c_{j}} \in N({b_i})} {u_{{c_{j}} \to {b_i}}}.
%\end{equation}
%\hspace{0.3cm}
%The recovery operator can be chosen

%\hspace{0.3cm}
\subsection{Quantum Sum-Product Decoder over the Depolarizing Channel}
In this section, we investigate the dependence of the performance of a quantum LDPC code on the estimated flip probability $\hat{f}$ of a depolarizing channel using  the same quantum LDPC code simulated in \cite{ShaneACTW2012}, which is Code~A of \cite{Tan:2010}. In each decoding process, the decoder performed an iterative message passing algorithm (sum-product decoding algorithm) until it either found a valid codeword (regardless of whether it is the word transmitted) or reached a maximum number of $200$ iterations. The simulation plots herein is the probability of block error ($P_{BLER}$) as a function of the estimated flip probability.

%where the block error probability is defined
%\begin{equation*}
%BLER = \frac{{\text{Number of failures}}}{{\text{Total number of runs} }}.
%\end{equation*}

\begin{figure}[htp]
\centering
\includegraphics[width=3.4in]{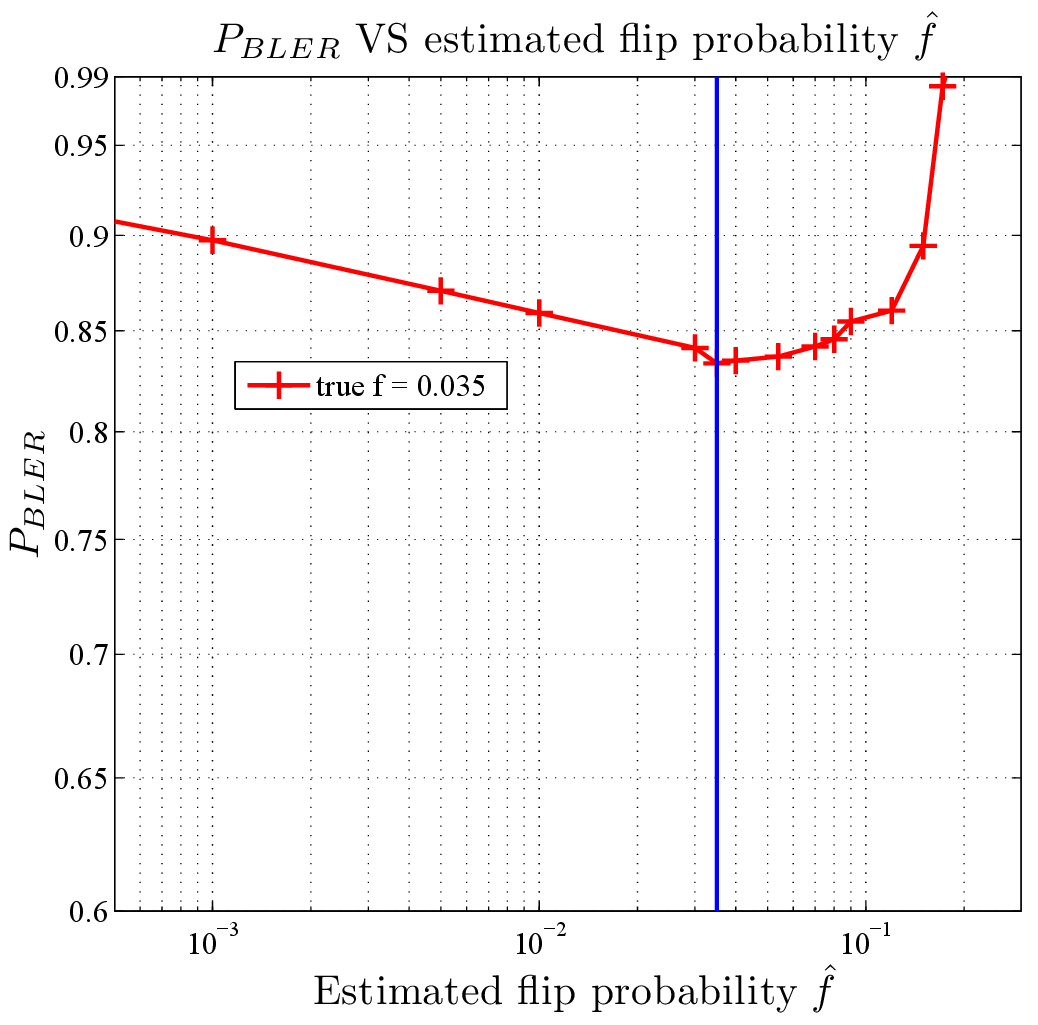}
\vspace{-0.3cm}
\caption{\hspace{0.1cm}Probability of block error as a function of estimated flip probability when the true flip probability is fixed.}
\label{fig:fvsfhat}
\end{figure}

In the simulations, the noise vectors were generated to have weight exactly $fN$, where $N$ was the block length of the code $(N = 1034)$ and $f$ is the true flip probability for the depolarizing channel. The decoder assumed an estimated flip probability $\hat{f}$. We varied the value of $\hat{f}$ while the the true flip probability $f$ is fixed. The results of our simulations are shown in Fig. \ref{fig:fvsfhat}.

Similar to the case of classical LDPC codes discussed earlier, we can see from Fig. \ref{fig:fvsfhat} that optimal performance in the quantum LDPC code can be obtained when the input at the decoder is the true flip probability, \emph{i.e} exact channel knowledge known. The trend of the curve in Fig. \ref{fig:fvsfhat} also shows an overestimate of $f$ is less costly than an underestimate of $f$, provided that the estimation of channel flip probability, $\hat{f}$, is less than some threshold $f^{Q}_{thr}$. For the code shown in Fig. \ref{fig:fvsfhat}, the theoretical threshold is $f^{Q}_{thr}=0.1893$ (equivalent to a classical rate $1/2$ code over $4$-ary symmetric channel). If $\hat{f}>f^{Q}_{thr}$, there is a catastrophic increase in the error probability. In the following section, we show that an improvement in performance of the sum-product decoder can be achieved if $\hat{f}<f^{Q}_{thr}$.
%that arising from an underestimation of the channel flip probability. When the estimated flip probability
%reaches a limit (beyond the limit of $0.189$ for a classical rate $3/4$ code over $4$-ary symmetric channel),
%there is a catastrophic increase in the error probability. This result indicates that in the quantum case
%(just as found in the classical case), there also exist a limit to the estimated channel flip probability.
%Since the value of this limit is known as the threshold (we denote the threshold value for quantum case as $f^{Q}_{thr}$) of noise level that the
%code would guarantee reliable transmission. Therefore, any estimation of the channel beyond this threshold
%is meaningless. In the following, we investigate the possibility of improve the performance of the
%sum-product decoder when this threshold value is considered.

\hspace{0.3cm}\hspace{0.3cm}
\subsection{Improved Decoding of Channel}

In this section, a numerical approach to improving the performance of the sum-product decoder is described. The asymmetric behavior of the sum-product decoder shown in Fig. \ref{fig:fvsfhat} implies that in the case of channel mismatch, an overestimation of the channel flip probability is more desirable than underestimation.

Consider  the case where a decoder can only attain partial channel information by probing the quantum
channel using un-entangled or entangled quantum states.
Given such partial information we will then weight our estimate of the channel parameter (at the decoder
side) to larger values (rather than smaller values) of the estimated flip probability.
%A schematic of our new decoding strategy is shown in Fig. \ref{fig:scheme}.

%\begin{figure}[htp]
%\centering
%\includegraphics[width=3.6in]{./Figures/scheme.eps}
%%\includegraphics[width=3.5in]{./scheme.eps}
%\vspace{-0.3cm}
%\caption{\hspace{0.1cm}An outline of the quantum decoder strategy for misidentified depolarizing channels.}
%\label{fig:scheme}
%\end{figure}

%\begin{figure}[htp]
%\centering
%\includegraphics[width=3.0in]{./Figures/scheme.eps}
%%\includegraphics[width=3.3in]{./scheme.eps}
%\caption{\hspace{0.1cm}An outline of the quantum decoder strategy for misidentified depolarizing channels.}
%\label{fig:scheme}
%\end{figure}

For a given true flip probability $f$, the probability of block error shown in Fig. \ref{fig:fvsfhat} can be fit approximately by:

\begin{equation}
\centering
\label{equ:fitfunction}
P_{BLER}^{(f)}(\hat{f}) \approx a + b\hat{f}^{3} + c\hat{f}^{5} + d\hat{f}^{7} + e\sqrt{\hat{f}} ln(\hat{f}),
\end{equation}
where $a,b,c,d,e$ are constants (the approximation gives a $2\%$ tolerance).
Assuming  our estimator of  $\hat{f}$ is centered on the true flip probability (\emph{i.e.} an unbiased estimator),
has a variance derived from its quantum Fisher information (\emph{i.e.} an optimal estimator),  and has a known probability density function $P(\hat{f})$,
we can then make an estimate of what constant should be added to any estimated $\hat{f}$ in order to maximally improve the decoder performance.
%that is

%\begin{equation}
%\centering
%\label{equ:gaussian}
%P(\hat{f}) = \frac{1}{{\sqrt {2\pi } \sigma }}{e^{ - \frac{{{{(\hat f - \mu)}^2}}}{{2{\sigma ^2}}}}}
%\end{equation}
%characterized by mean a $\mu$ and variance $\sigma^2$.
Note that, for the case where the qubit probe is in an unentangled state, the quantum Fisher information about $f$ can be shown to be $(N_m J(f))^{-1} = [f(2-f)]$. The average probability of block error for a given $f$ can then be estimated using the equation
\begin{equation}
\label{equ:average}
\centering
\tilde{P}_{BLER}^{(f)} = \int\limits_{0}^{f^{Q}_{thr}} {P(\hat f)} P_{BLER}^{(f)}(\hat f)d\hat{f}.
\end{equation}
%where $f^{Q}_{thr}$ denotes the theoretical limit of estimated flip probability.
The performance of the sum-product decoder can be improved if a factor $\Delta \hat{f}$ is added to the estimated value of $\hat{f}$. That is,  $\hat{f} \to \hat{f}+\Delta \hat{f}$.  The question then becomes, given some channel what is the optimal $\Delta \hat{f}$ that minimizes the expected probability of error? To answer this, equation (\ref{equ:average}) is modified to
\begin{equation}
\label{equ:averageModified}
\centering
\tilde{P}_{BLER}^{(f)}(\Delta \hat{f}) = \int\limits_{0}^{f^{Q}_{thr}} {P(\hat f)} P_{BLER}^{(f)}(\hat f + \Delta \hat{f})d\hat{f},
\end{equation}
The optimal $\Delta \hat{f}$ is then the solution to
\begin{equation}
\label{equ:deltafEquation}
\centering
\frac{\partial }{{\partial \Delta \hat{f}}}{{\tilde P}_{BLER}^{(f)}}(\Delta \hat{f}) = 0.
\end{equation}
One could repeat this process for a range of true channel flip probabilities, and derive an estimate of the $\Delta \hat{f}$ averaged over the range of  true flip probabilities where QECC can be expected to be of relevance, that is
\begin{equation}
\label{equ:averageDf}
\centering
\Delta \hat {f}_{avg} = \int\limits_0^{f_{thr}^Q} {P(f|\hat f)\Delta {{\hat f}^{(f)}}df}.
\end{equation}

\begin{figure}[htp]
\centering
\includegraphics[width=3.5in]{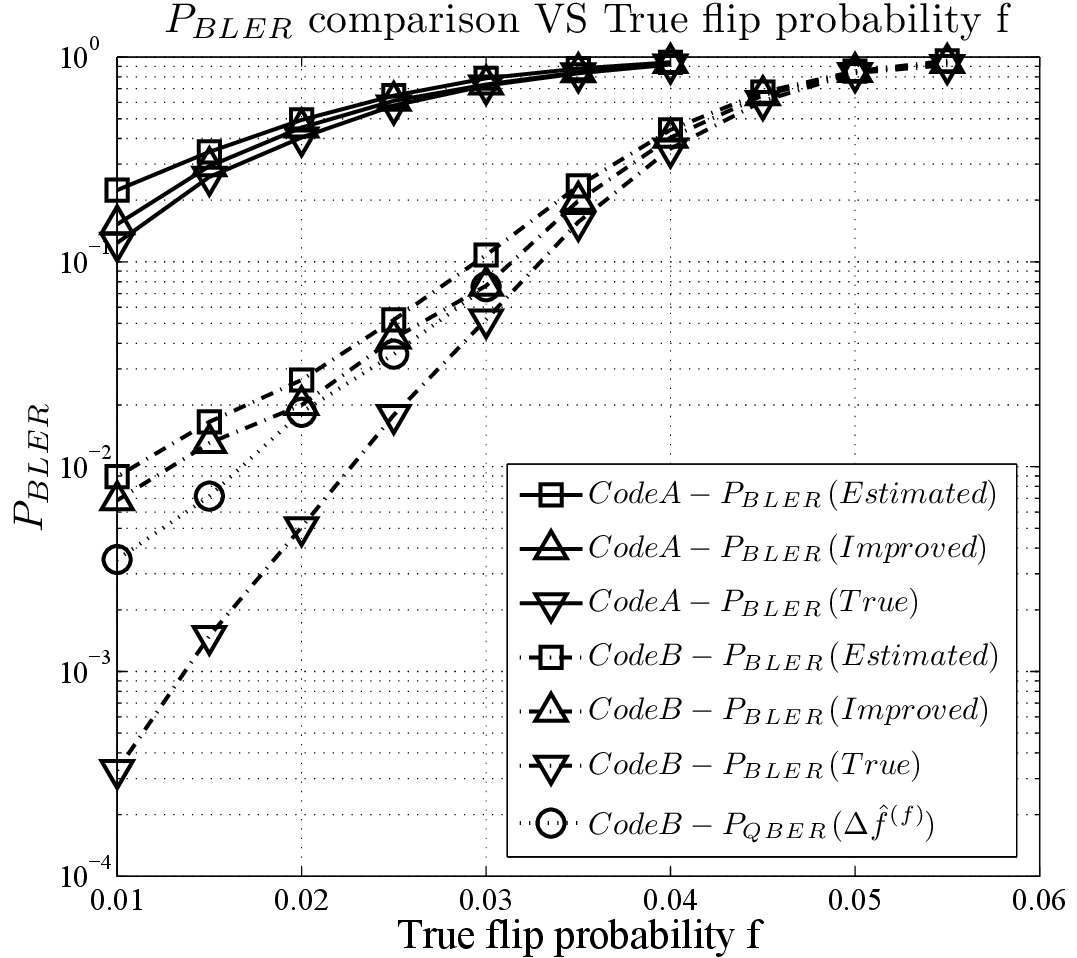}
\vspace{-0.3cm}
\caption{\hspace{0.1cm}Comparison of $P_{BLER}$.}
\label{fig:improvedBLER}
\end{figure}

\begin{figure}[htp]
\centering
\includegraphics[width=3.5in]{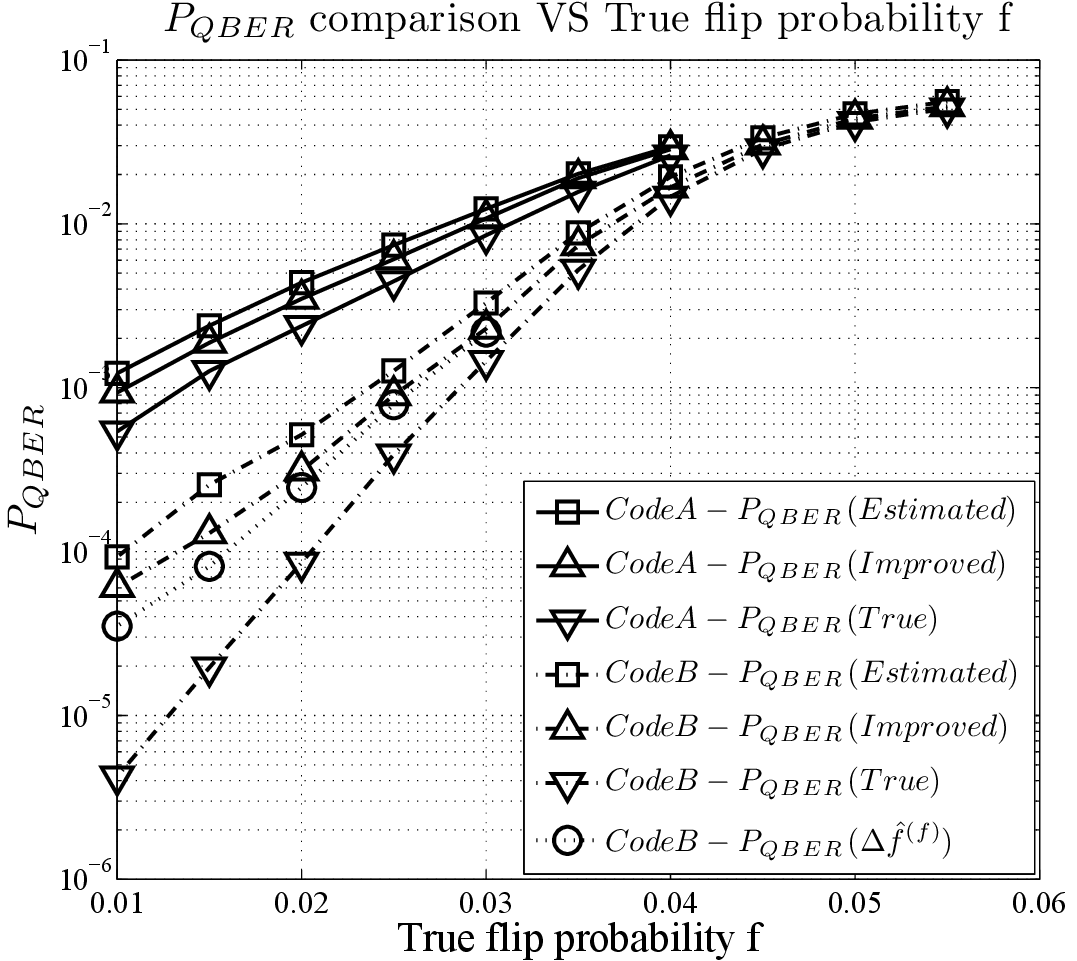}
\vspace{-0.3cm}
\caption{\hspace{0.1cm}Comparison of $P_{QBER}$.}
\label{fig:improvedQBER}
\end{figure}

%Assuming equation (\ref{equ:gaussian}) and channels with true flip probability $f$, use the code assumed for Fig. \ref{fig:fvsfhat}. Based on the method above, a set of different true flip probability ($f$) is simulated. The corresponding optimal $\Delta \hat{f}$ is computed numerically and listed in TABLE \ref{table:tvsd}.
For the same code (Code A) as that used in Fig. (\ref{fig:fvsfhat}), assume a uniform distribution for $P(f|\hat{f})$, and taking $N_m=1$ in the Fisher information, we found that value of $\Delta \hat {f}_{avg}$ to be very weakly dependent on $f$ (see Table \ref{table:tvsd} ). This means that simply adding to each estimated  $\hat{f}$  the additional factor  $\Delta \hat {f}_{avg}$ led to substantial performance improvement. The magnitude of this improvement can be seen in Fig. \ref{fig:improvedBLER} and \ref{fig:improvedQBER}. In these figures $\Delta \hat {f}_{avg} \approx 0.0025$ is applied at the
the decoder to provide the improved error correction (shown are the fraction of blocks in error $P_{BLER}$, and the fraction of qubits in error $P_{QBER}$), denoted as ``Improved''.
The notation ``Estimated'' in these figures is for the case where the input to the sum-product decoder
is $\hat{f}$ only, whereas the notation ``True'' is for the case where the input to the decoder is
the true flip probability $f$. As can be seen improvements of up to $\sim 50\%$ can found from the new strategy (Improved), relative to the case of just utilizing the estimated $\hat{f}$. Similar results to those shown  were found for other codes investigated, although the factor to be added was found to be a function of the code. For example, in another code investigated (Code B using Construction method III of \cite{Tan:2010} block size 1032) a  $\Delta \hat{f}_{avg} \approx 0.0032$ was found to be better (the corresponding optimal $\Delta \hat{f}$ for each different true flip probability $f$ for Code B is also listed in TABLE \ref{table:tvsd} and see also Fig. \ref{fig:improvedBLER} and \ref{fig:improvedQBER} for simulation improvement). Of course, improved channel estimation also alters the details of our analysis, with more accurate measurements (\emph{e.g}. a higher number of measurements $N_m$ of the channel) leading to smaller $\Delta \hat {f}_{avg}$, and smaller improvements in performance. This highlights the trade-off between channel estimation and QECC.

\begin{table}[h]
\centering

\begin{tabular}{|c|c||c|c|}
\hline
\multicolumn{2}{|c|}{Code A} & \multicolumn{2}{|c|}{Code B}\\
\hline
$f$ & $\Delta \hat{f}$ & $f$ & $\Delta \hat{f}$ \\ \hline
0.04  &  0.00332    & 0.05       & 0.00376 \\ \hline
0.035 &  0.00286    & 0.045      & 0.00349\\ \hline
0.03  &  0.00258    & 0.04       & 0.00271\\ \hline
0.025 &  0.00197    & 0.035      & 0.00245 \\ \hline
0.02  &  0.00255    & 0.03       & 0.00271 \\ \hline
0.015 &  0.00271    & \null     & \null \\ \hline
0.01  &  0.00152    & \null     & \null \\ \hline
\end{tabular}
\caption{\hspace{0.1cm}Optimal $\Delta \hat{f}$ for different true flip probability.}
\label{table:tvsd}
\end{table}

Finally, It is perhaps worth illustrating how the use of $\Delta \hat{f}_{avg}$, rather than the optimal $\Delta \hat{f}^{(f)}$, impact the results.
If optimal $\Delta \hat{f}^{(f)}$ for each true $f$ is applied for $f<0.025$, the error performance
is further improved (circle dashed line depicted in Fig. \ref{fig:improvedBLER} and \ref{fig:improvedQBER}).

%The result in these plots showed that both probabilities of $P_{BLER}$ and $P_{QBER}$ for the improved case are reduced, and approaches to the performance curve denoted by ``True''.
%It is worth to mention that our improved results obtained by applying the proposed methodology does not generalize all the quantum LDPC codes.
%We further examine the behavior of the decoder for another quantum LDPC codes based on construction III given in \cite{Tan:2010}. The constructed code graph has the same dimension and degrees as the one simulated above.
%The asymmetric behavior of the decoder preserved, while the performance is different compare to the performance simulated above. One reason is that the code simulated here is conventionally known as a `bad' code because it contains the cycle of length $4$, which degrades the performance of the sum-product decoder significantly.
\section{Conclusion}

In this work we have investigated possible improvements in the decoding strategies of quantum LDPC decoders in the quantum depolarization channel. The importance of the channel mismatch effect in determining the performance of quantum LDPC codes has very recently been shown to lead to a degradation in the qubit error performance.  In this work we have illustrated how such a performance  gap in the qubit error performance can be substantially reduced. The new strategies for quantum LDPC decoding we provided here are based on previous insights from classical LDPC decoders in  mismatched channels, where an asymmetry in performance is known as a function of the estimated bit-flip probability. We first showed  how similar asymmetries  carry over to the quantum depolarizing channel. We then showed  that when a weighted estimate of  the depolarization flip parameter to larger values is assumed,  performance improvement by as much as $50\%$ was found. We conjecture that all quantum channels which are misidentified, or for which only partial channel information is available, will benefit from  similar decoding strategies to those outlined here.

The work outlined here will be of practical importance when large-scale quantum networks are built, and sophisticated quantum error correction codes are deployed in order to maintain the entanglement between the distributed entangled qubit pairs that underpins these emerging networks. The strategies described here will ultimately manifest themselves in an improved performance of entanglement-based QKD, or any other entanglement-based quantum communication application, deployed over such future networks.
%Again we believe that such entanglement-assisted codes in the misidentified channel will likely benefit from the decoding strategies we outlined here.

\section*{Acknowledgments}
This work
has been supported by the University of New
South Wales, and the Australian
Research Council (ARC).

%\end{thebibliography}

% that's all folks
\end{document}